\documentclass[prb,twocolumn,showpacs,preprintnumbers,floatfix,english]{revtex4}

\usepackage[utf8]{inputenc}
\usepackage[T1]{fontenc}
\usepackage{babel}

\usepackage{ifpdf}
\usepackage{graphicx}
\usepackage{color}
\usepackage{lineno}
\usepackage{pslatex}

\usepackage{bm}
\usepackage{dcolumn}

\usepackage{amsmath}
\usepackage{amssymb}
\usepackage{amscd}

\ifpdf
\usepackage{epstopdf}
\usepackage[%
  pdftitle={lGe2Se3},%
  pdfauthor={},%
  pdfsubject={lGe2Se3},%
  pdfkeywords={lGe2Se3},%
  pdfstartview=FitH,%
  bookmarks=true,%
  bookmarksopen=true,%
  breaklinks=true,%
  colorlinks=true,%
  linkcolor=blue,anchorcolor=blue,%
  citecolor=blue,filecolor=blue,%
  menucolor=blue,pagecolor=blue,%
  urlcolor=blue]{hyperref}
\else
\usepackage[%
  pdftitle={lGe2Se3},%
  pdfauthor={},%
  pdfsubject={lGe2Se3},%
  pdfkeywords={lGe3Se3},%
  breaklinks=true,%
  colorlinks=true,%
  linkcolor=blue,anchorcolor=blue,%
  citecolor=blue,filecolor=blue,%
  menucolor=blue,pagecolor=blue,%
  urlcolor=blue]{hyperref}
\fi

\include{bibsymb}

\begin{document}
\draft

\title{Crucial effect of glass processing and melt homogenization on the fragility of non-stoichiometric chalcogenides}

\author{Sriram Ravindren$^1$, K. Gunasekera$^1$, Z. Tucker$^1$, A. Diebold$^2$, P. Boolchand$^1$, M. Micoulaut$^3$}

\affiliation{$^1$School of Electronics and Computing Systems, College of Engineering and Applied Science, University of Cincinnati
Cincinnati, OH 45221-0030, USA}
\affiliation{$^2$Department of Physics, University of Cincinnati, Cincinnati, OH 45221-0011, USA}
\affiliation{$^3$Laboratoire de Physique Th\'{e}orique de la Mati\`ere Condensée,
Universit\'{e} Pierre et Marie Curie, 4 Place Jussieu, F-75252 Paris Cedex 05, France}

\date{\today}

\begin{abstract}
The kinetics of homogenization of binary As$_x$Se$_{100-x}$ melts in the As concentration range 0\% $<$ $x$ $<$ 50\%
are followed in FT-Raman profiling experiments, and show that 2 gram sized melts in the middle
concentration range 20\% $<$ $x$ $<$ 30\% take nearly two weeks to homogenize when starting materials are
reacted at 700$^o$C. In glasses of proven homogeneity, we find molar volumes to vary non-monotonically
with composition, and the fragility index ${\cal M}$ displays a broad global minimum in the 20\% $<$ $x$ $<$ 30\% range of
$x$ wherein ${\cal M}<$ 20. We show that properly homogenized
samples have a lower measured fragility when compared to larger under-reacted melts. 
The enthalpy of relaxation at T$_g$, $\Delta$H$_{nr}$($x$) shows a minimum in the 27\% $<$ $x$ $<$ 37\%
range. The super-strong nature of melt compositions in the 20\% $<$ $x$ $<$ 30\% range suppresses melt
diffusion at high temperatures leading to the slow kinetics of melt homogenization. 
\end{abstract}


\maketitle

\section{Introduction}
Fragility of glass forming melt historically emerged from investigations of the temperature dependence
of viscosity\cite{r1}. The manner in which viscosity ($\eta$) of such melts increases as
melts are cooled to the glass transition temperature $T_g$ can be used to characterize viscous slow down of supercooled liquids.
Since $\eta$ is proportional to a shear relaxation time $\tau$ through the Maxwell relation ($\eta$=G$_\infty \tau$) where G$_\infty$ is the bulk modulus at infite frequency, it is convenient to define the fragility ${\cal M}$ index as
\begin{eqnarray}
\label{frag}
{\cal M}\equiv \biggr[{\frac {d\log_{10}\tau}{dT_g/T}}\biggr]_{T=T_g}
\end{eqnarray}
The dimensionless slope ${\cal M}$, examined in a wide variety of glass forming liquids\cite{r2}, is found to vary between
a high value\cite{simon} of 214 to a low value\cite{r13} of 14.8. Melts possessing a high value are termed as “{\em fragile}” while the ones with
a low value are "{\em strong}". Fragile melts usually show a highly non-exponential variation of $\eta$(T) while
strong ones have a nearly Arrhenius variation\cite{r2}. At high T, both variations lead to the same value of $\eta$\cite{r3}.
The non-exponentiality of $\eta$ in glass forming melts has been fitted in terms of various functions such
as Vogel-Fulcher-Tamman (VFT\cite{r4}) given by:
\begin{eqnarray}
\label{vft0}
\eta=\eta_0\exp\biggl[{\frac {A}{T-T_0}}\biggr]
\end{eqnarray}
where $A$ and $T_0$ are fitting parameters, or the Mauro-Yue-Ellison-Gupta-Allan (MYEGA\cite{r5}) or Avramov-Milchev (AM\cite{r6}) functional forms.
\par
Since the inception of the strong-fragile melt classification, $\eta$(T) variation in stoichiometric glass forming
melts such as the oxides, chalcogenides, alcohols, sugars, and organic polymers has been well
documented\cite{r7} in a plot initially introduced by Laughlin and Uhlmann\cite{r8}, and subsequently popularized by Angell\cite{r9}. 
The case of non-stoichiometric inorganic glass forming melts
has been explored much less. In an early study on the As-Ge-Se ternary\cite{r10} it was recognized
that the fragility index of such alloyed melts shows a minimum near a mean coordination number $\bar r$ =2.40, identified with the location of a 
mean-field flexible to rigid transition\cite{r11,r12}. 
More recent work on the Ge$_x$Se$_{100-x}$ binary has shown\cite{r13} that the fragility index ${\cal M}$ 
takes on a rather low value of 14.8(5) near a Ge concentration of $x$ = 22\% corresponding to a mean coordination
number $\bar r$ = 2.44 residing in the Intermediate Phase of corresponding bulk glasses\cite{r14,r15,r16} where the enthalpic relaxation 
at the glass transition temperature is minuscule. A theoretical link between enthalpic changes, fragility and isostatic character of the glass network 
has been established from a simple Keating model reproducing the behavior of covalent glass-forming liquids\cite{r17}.
\par
The observation of important fragility minima (i.e. super-strong melts) has actually profound consequences. The super-strong character over selected 
compositional ranges has, in fact, deep implications on the homogenization of corresponding melts. This is the case because melt
viscosity for such compositions at high temperatures, where melts are equilibrated, far exceed those of the more fragile compositions outside of this
compositional window. The wide variations in melt viscosity with composition thus leads -in a natural fashion- to hindered diffusion which is manifested as a slow batch homogenization\cite{r17b,r17t}.
\par
In this work we show, for the first time to the best of our knowledge, that As$_x$Se$_{100-x}$ melts in the 20\% 
$<$ $x$ $<$ 35\% composition range are quite strong with a very low fragility index ${\cal M}$ $<$ 20, and the lowest one being found for the $x$=24\% composition, which corresponds to a fragility index of ${\cal M}$=16.4, i.e. slightly less than that of silica, which is traditionally thought of as the archetypical strong liquid\cite{r18,r19,r20}. Melt compositions outside this broad compositional window possess higher fragility indices. 
This leads amongst other things to the dynamics of melt homogenization during glass melting to qualitatively slow down. Melt homogenization is a critical subject that is, unfortunately, often overlooked in the literature. Here we bring a quantitative connection between fragility measurements and the ease of homogenization.  
Several broad consequences emerge. One, these conditions, as in the
case of the Ge-Se binary\cite{r13}, also cause As-Se melts to undergo slow homogenization as we demonstrate
here directly from FT-Raman profiling experiments. Second, the broad fragility window residing in the
mean-coordination number range of 2.20 $<$ $\bar r$ $<$ 2.35, correlates more or less with a minimum in the enthalpy of
relaxation at $T_g$ in corresponding glasses (reversibility window\cite{r21}), and both are shifted to $\bar r$ $<$ 2.40. Third,
because of slow homogenization of As-Se melts, physical properties of these melts/glasses are notoriously 
non-reproducible\cite{r22,r23}. We highlight this important issue for molar volumes and fragility index
measurements in this context. Previous conclusions drawn for Ge-Se are fully recovered\cite{r13} and underscore the generality 
of the above statements.
\par
The
paper is organized as follows. After presenting the experimental results in section \ref{exp}, and the method of Raman profiling which allows to check for the homogeneity of the samples, we present the experimental results on fragility activation energy for relaxation and molar volume in section \ref{results}. We discuss the consequences of the findings in section \ref{cons}, and show that general correlations can be drawn. Finally, the conclusions are summarized in section \ref{conclusion}.
\begin{figure}
\includegraphics*[width=0.8\linewidth]{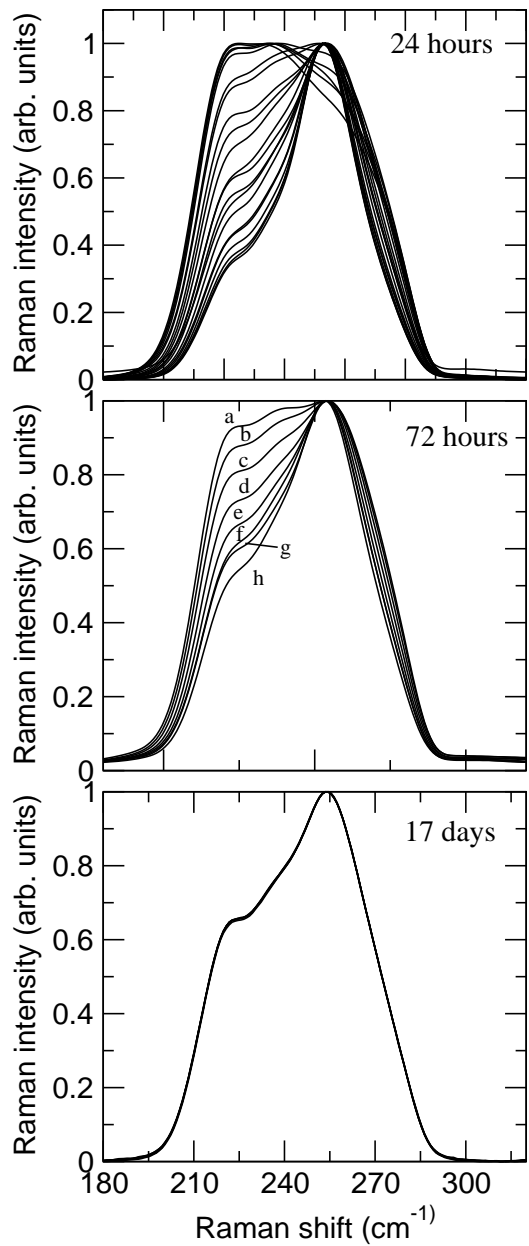}
\caption{\label{Raman_prof} FT Raman profiling of a As$_x$Se$_{100-x}$ melts at x = 24\% taken after indicated reaction times T$_R$ (24 hours, 72 hours and 17 days). 
Melts are heterogeneous at T$_R$ $<$ 24 hours, but steadily homogenize as the reaction time increases to several days. We show that the
underlying factor is the fragility of melts.
}
\end{figure}
\section{\label{exp} Experimental}
\subsection{Synthesis of glasses}
Bulk glasses of the As$_x$Se$_{100-x}$ binary were synthesized using elemental Se (5N purity, 3-4~mm dia.,
Cerac Inc.), and As$_2$Se$_3$ (5N purity, 1-6~mm granules, Noah Technologies) as precursors. As-rich
glasses ($x$ $>$ 40\%) made use of As$_2$Se$_3$ and As$_{50}$Se$_{50}$; the latter synthesized using pure 
As (5N purity, large lumps, Cerac Inc.) and As$_2$Se$_3$ precursors. Prior to their use, the clean ampoules were dried in a
vacuum oven at 90$^o$C for 24 hours. The ingredients were mixed in the desired ratio by weight to give
a total of 2~grams (to an accuracy of 0.1~mg) per sample and subsequently batched in quartz tubing (5~mm ID, 
7~mm OD). The precursors were not crushed in order to keep to a minimum the surface area
available to oxidation and water adsorption. It is unclear if As-Se melt homogenization is
significantly accelerated by crushing the precursors; factors such as viscosity overwhelmingly
determine the homogenization period. The mostly nodular precursors served to facilitate
attainment of high vacuum ($\simeq$3$\times$10$^{-7}$~ Torr) using a liquid N$_2$-trapped high vacuum 
pumping system, and the quartz tubes sealed using a Hydrogen-Oxygen torch. Typical ampoule length was between
75-90~mm. The sealed ampoules, positioned vertically in a T-programmable vertical tube furnace,
were reacted by initially ramping T up linearly from 25$^o$C to 700$^o$C over 13 hours, and then held at
700$^o$C until melts homogenized for periods lasting up to two weeks in some cases. Prior to removing
the samples from the furnace, the temperature was lowered to 50$^o$C above the reported liquidus\cite{r25,r26}
and equilibrated at that T for 2 hours, prior to a water quench.
\par
Melts were periodically FT–Raman profiled (see below) to map the diminishing heterogeneity. Once
melt homogeneity was established, these were $T_g$-cycled {\em in situ} by holding isothermally at T$_g$($x$)+20$^o$C
for 30 minutes (to remove residual stresses) in a box furnace, and slow cooled at 3$^o$C/min to room
temperature to realize homogeneous bulk glasses. Glasses were then removed from evacuated quartz
tubes, and weighed for Modulated-DSC ($\simeq$ 20~mg), Fragility ($\simeq$10~mg) and Molar volume ($>$200~mg)
measurements.

\subsection{Raman line profiling of quenched melts}

\begin{figure}
\includegraphics*[width=0.9\linewidth]{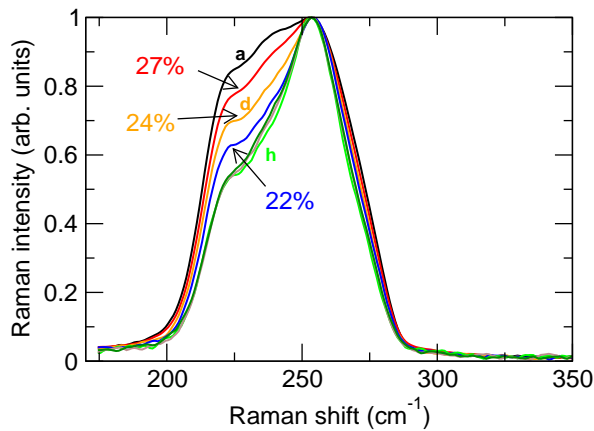}
\caption{\label{fig2} Fractionation of the As$_{24}$Se$_{76}$ melt reacted at 700$^o$C for 3 days is revealed in this stacked Raman
spectra. Some of the observed lineshapes are characteristic of specific compositions. This plot illustrates the extent to which the melt resists homogeneity with a spread of nearly 7\%.
}
\end{figure}
Melt homogeneity was ascertained using an FT-Raman spectroscopic line profiling method\cite{r17b,r17t}. 
Raman vibrational density of states (VDOS) of As$_x$Se$_{100-x}$ glasses vary measurably with glass
compositions ‘$x$’. Melt heterogeneity was tracked by recording FT-Raman spectra along the length of the
melt column encased in quartz tubing, first after 72 hours of reaction, and then every 48 hours
thereafter until all lineshapes recorded along the melt column became identical when they were considered to be fully homogeneous.
We used a Thermo Nicolet NXR FT-Raman module with 1064~nm (1.17~eV) radiation from an Nd-YAG
laser to excite the scattering. The laser spot size was at 50~$\mu$m. A typical measurement at a single
location involved 100 scans at 2~cm$^{-1}$ resolution (219.70 seconds/location) using a laser power of 105
mWatts.
\par
Raman spectral acquisition was programmed for typically 8 equidistant sample-focused points along the
length of the quartz tube. These spectra were baseline adjusted, amplitude corrected and stacked to
qualify spectral overlap. The uniqueness of a composition’s spectral signature implies that complete
spectral overlap is critical to establishing sample homogeneity. Observable differences in stacked
lineshapes suggest variations in local sample stoichiometry ‘$x$’, implying heterogeneity of the reacted
batch composition. After full homogenization was established, a library of high resolution Raman
spectra was acquired at each composition in the 0\% $<$ $x$ $<$ 50\% range. The library provided a basis to
track the evolution of homogeneity over time.
\par
Fig. \ref{Raman_prof} illustrates an example of a Raman profiled result. The three panels provide a global view of melt
homogenization kinetics at x = 24\%. It is observed that all 8 Raman lineshapes taken along the melt
column coalesce after heating at 700$^o$C for 17 days. From such studies we
were able to generate a library of Raman spectra of homogenized melts at 20 compositions in the range
0\% $<$ $x$ $<$ 50\%, and it is used in analyzing the profiled Raman spectra given that important information on 
the nature of structural heterogeneity across a melt column is contained in the
observed spread of Raman line-shapes as highlighted by profiling
results of a melt at x = 24\% after being reacted at 700$^o$C for 3 days (Fig. \ref{fig2}). 
It is seen that three of the observed line-shapes in that group of 8 correspond
exactly to melt stoichiometry of x = 22\%, 24\% and 27\% as indicated in Fig. \ref{fig2}. These data show that the As
content along the length column exhibits a gradient; it has a maximum value of about x = 28\% at the
bottom and a minimum value of x = 21\% at the top of the melt column. Thus, we find that for the 2~gram
sized melt, reacted at 700$^o$C for 3 days, the As stoichiometry varies by nearly 7\% across the melt column, and would inevitably lead
to some spread if a physical measurements were to be undertaken from glasses having been prepared at this reaction time.
\begin{figure}[t]
\includegraphics*[width=\linewidth]{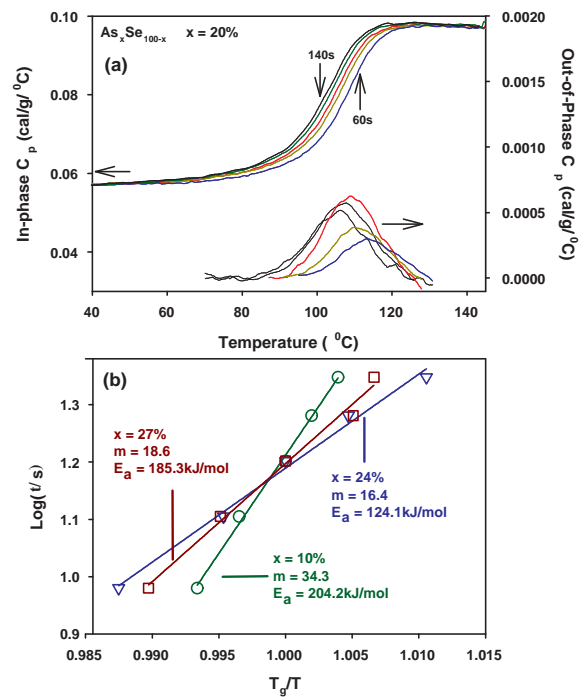}
\caption{\label{cp} The real (a) and imaginary (b) parts of the complex heat capacity C$_p^*$ for an As$_{20}$Se$_{80}$ glass
were obtained from MDSC measurements at a rate of 3$^o$C/min and a modulation amplitude of 1$^o$C over
five periods (1$^o$C/60s, 80s, 100s, 120s, 140s). The fragility index was extracted from the slope of the
curves at $T_g/T$=1 in the Angell plot.
}
\end{figure}
\section{\label{results} Results}
\subsection{Fragility measurements}
Melt fragility were established by recording the complex C$_p$ heat flow (C$_p^{real}$, C$_p^{imaginary}$) 
near $T_g$ using a TA Instruments Q2000 MDSC system. A 10~mg quantity of a glass sample sealed in an aluminum
pan, was cooled starting from $T_g$+50$^o$C across the glass transition temperature and then heated back to
$T_g$+50$^o$C. Such experiments were undertaken at 5 modulation periods (60s, 80s, 100s, 120s, and 140s).
\par
\begin{figure}
\includegraphics*[width=0.9\linewidth]{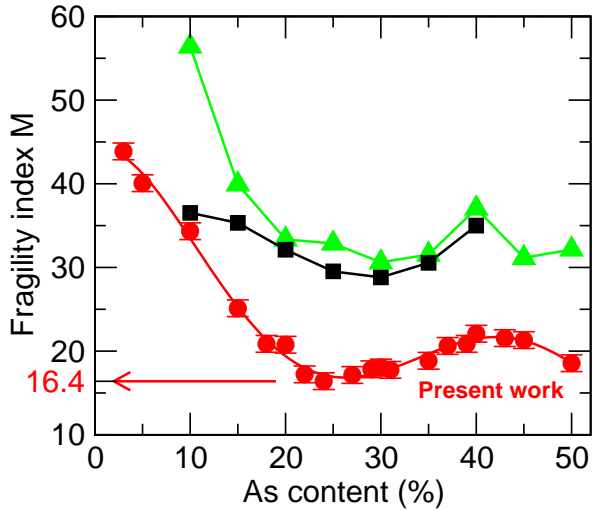}
\caption{\label{fragile} Compositional variation of the fragility index ${\cal M}$ of the 
As-Se system as reported in present work ($\color{red}\bullet$), compared with other recent reports ($\blacksquare$, Ref. \onlinecite{r22}; $\color{green}\blacktriangle$, Ref. \onlinecite{r23}). 
}
\end{figure}
Fig. \ref{cp} shows the C$_p^{real}$ or in-phase C$_p$ to display a rounded step with the step shifting to higher T as the
frequency increases. The C$_p^{imaginary}$ term or out-of phase C$_p$ displays a Gaussian-like peak and the peak
steadily shifts to higher T as the modulation frequency increases.
\par
In these scans, the peak location of the C$_p^{imaginary}$ term recorded at 100~sec modulation period defines the
glass transition temperature $T_g$. At the peak location, the relaxation is defined by the condition
expressed by $\omega\tau=1$, where $\tau$ is the relaxation time at the peak and $\omega$ is the relaxation frequency.
Thus, the shift of the peak in C$_p^{imaginary}$ to higher temperatures as the modulation frequency is increased
signals a reduction in the melt shear relaxation time. By plotting the variation of $\log_{10}\tau$ against $1/T$
normalized to $1/T_g$, we deduced the fragility index ${\cal M}$ using equation (\ref{frag}). Fig \ref{cp}b illustrates 
the results for the case of 3 melt compositions of x = 10\%, 24\% and 27\%. The enthalpic activation energy $E_a$ associated
with the fragility ${\cal M}$ is related to $T_g$ by the following relation:
\begin{eqnarray}
E_a={\cal M}RT_g\ln{10}
\end{eqnarray}

\begin{figure}
\includegraphics*[width=0.9\linewidth]{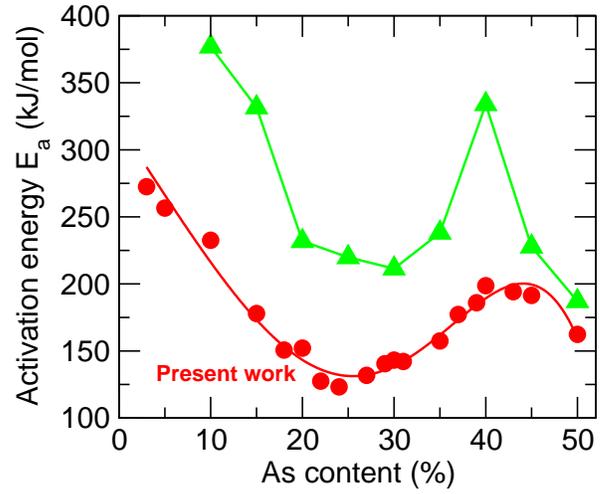}
\caption{\label{activ} Compositional variation of the activation energy $E_a$ ($\color{red}\bullet$, present work), computed from the fragility index (Fig. \ref{fragile}), and compared to previous results from Yang et al. ($\color{green}\blacktriangle$, Ref. \onlinecite{r23}).  
}
\end{figure}

The compositional variation of the fragility index ${\cal M}$($x$) and the activation energy $E_a$($x$) for our As$_x$Se$_{100-x}$
melts of proven homogeneity are displayed in Fig. \ref{fragile} and \ref{activ}, respectively. Our results show both 
${\cal M}$($x$) and $E_a$($x$) to reveal a broad global minimum centered around x = 25\%. Specifically, melt
compositions in the 20\% $<$ $x$ $<$ 35\% display a fragility index ${\cal M}$ of 20 or less. The lowest fragility index (16.4) is
obtained for a melt at $x$ = 24\%. The melts in this strong region homogenize far more slowly compared to
the more fragile melts, most of which homogenize in under 3 days without rocking. This correlation is
remarkable, and we comment on it below.
\par
In Fig \ref{fragile} we have also included fragility results from Viscosity measurements of Musgrave et al.\cite{r22} and
Complex C$_p$ measurements of G. Yang et al.\cite{r23}. There are clear similarities in global trends in all these data as all display a minimum for ${cal M}$ around the 25\% As content. The viscosity derived and present mDSC derived fragility index for a melt at x = 10\% are in excellent
agreement, although such is not the case at other melt compositions examined. The Complex C$_p$
derived fragility index of Yang et al.\cite{r23} are of interest because those were obtained the same measuring technique and the same 
calorimeter as in the present case. 

\subsection{Volumetric measurements}
Mass density of glassy solids provides useful characterization of a glassy solid. If one inverts it to deduce
the molar volume $V_m$, then one obtains information on network packing. Unlike scanning calorimetric
methods and Raman scattering that require a minuscule amount ($\simeq$20~mgms) of the glass, mass density
measurements were made using 200~mgms of a bulk glass to achieve an accuracy of 0.25\%. Thus, it is
possible to get a global average on a batch preparation and gauge its homogeneity.
\par
\begin{figure}
\includegraphics*[width=0.9\linewidth]{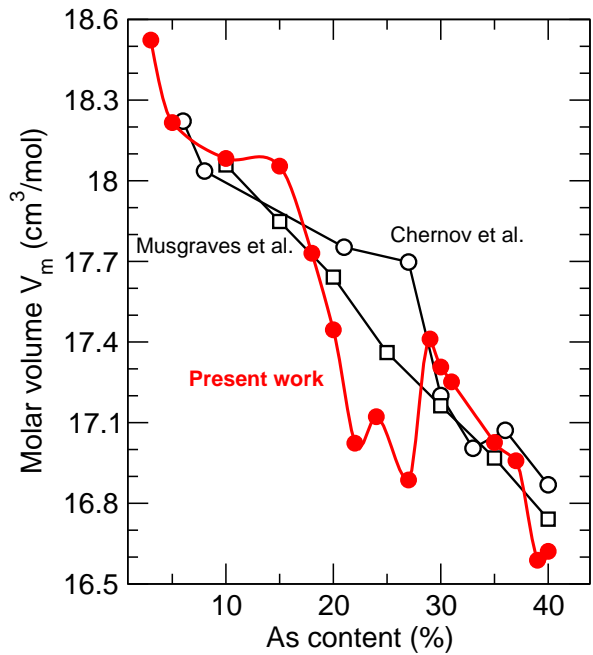}
\caption{\label{volume} 
Observed compositional molar volume profile V$_m$ in the As-Se system ($\color{red}\bullet$, present work), compared to previous measurements: 
($\square$, Ref. \onlinecite{r22}) and ($\circ$, Ref. \onlinecite{chernov}. The large variation in the reported values reflects the differences in melt homogenization in the samples.}
\end{figure}

We have performed mass density measurements on the bulk As$_x$Se$_{100-x}$ glasses using a quartz fiber and a
digital microbalance. A bulk glass specimen typically 200~mgm
in size was weighed in air and in 200 Proof Alcohol, and the density obtained using Archimedes principle.
A single crystal of Si was used to calibrate the density of alcohol and a single crystal of Ge used to check
the accuracy of density measurements. In Figure \ref{volume} we compare results of molar volume measurements
from the present work with several previous reports where complete trends across a wide range of
compositions are available.

\par
Several groups report\cite{r22,chernov,feltz} an almost linear variation of V$_m$($x$) with $x$ in the 0 $<$ $x$ $<$ 40\% range. 
In our case we observe a variation that departs from linearity particularly in the 20\% $<$ $x$ $<$ 30\% range. Nonlinear
variation of V$_m$($x$) have been reported for other systems as well\cite{germanate} and are related to the stress-free character of the network\cite{bourgel}. 
Given that the error in V$_m$($x$) in our measurements is twice the size of the data point, the variance between different group exceeds the
error of measurements and is most likely due to sample make up.
\section{\label{cons} Discussion}
The present experimental results raise several basic issues on the compositional dependence of physical properties of melts and glasses. In this section we discuss some of these.
\subsection{Fragility index and physics of melt homogenization}
We are not aware of any measurement of viscosity in As-Se at the reaction temperature of 700$^o$, and most of the available experimental data are being found either at much lower temperatures or are restricted to selected compositions\cite{r23,malek,rouxel}. Given the compositional trends in ${\cal M}$($x$) and the glass transition temperature T$_g$(x) accessible from the Complex C$_p$ measurements (Figs. \ref{cp}), we can calculate the variation of the viscosity $\eta$($x$) at 700$^o$C using two seminal models for viscosity variation with temperature. For the VFT, we obtain $\eta$($T$) using :
\begin{eqnarray}
\label{VFT}
\log_{10}\eta(T)=\log_{10}\eta_\infty+{\frac {(12-\log_{10}\eta_\infty)^2}{{\cal M}(T/T_g-1)+(12-\log_{10}\eta_\infty)}}
\end{eqnarray} 
which is an alternative form of equ. (\ref{vft0}), and for the MYEGA equations\cite{r5}:
\begin{eqnarray}
\label{MYEGA}
\log_{10}\eta(T)&=&\log_{10}\eta_\infty+(12-\log_{10}\eta_\infty){\frac {T_g}{T}}\times \\ \nonumber
& &\exp\biggl[\biggl({\frac {{\cal M}}{12-\log_{10}\eta_\infty}}-1\biggr)(T/T_g-1)\biggr]
\end{eqnarray} 
respectively.
\par
\begin{figure}
\includegraphics*[width=0.9\linewidth]{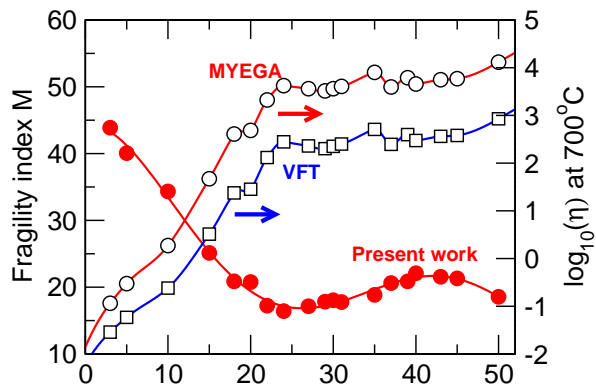}
\caption{\label{visco_myega} 
a) Left, fragility index in As-Se melts as a function of As content (same as Fig. \ref{fragile}). Right axis: Computed viscosity profile using the VFT and MYEGA equations at the reaction temperature of 700$^o$C.}
\end{figure}

It is instructive to consider a  model due to Mauro et al. (MYEGA), which gives $\eta$($T$) using equation (\ref{MYEGA}). At high $T$, and in fragile melts, there is evidence that the MYEGA model works better than the Vogel-Fulcher-Tamman (VFT) model or the Avramov-Milchev model. Using Maxwell’s relation $\eta_\infty=G_\infty\tau$ to obtain the viscosity, assuming an infinite frequency shear modulus $G_\infty\simeq$ 10~GPa for the As-Se system\cite{borisova}, we obtain $\eta$($x$) at the experimental reaction temperature (700$^o$C) for both the VFT and MYEGA equations, and these results are summarized in Figure \ref{visco_myega}.
\par 
A remarkable feature of the plot of Fig. \ref{visco_myega} is the six orders of magnitude change in viscosity of As$_x$Se$_{100-x}$ binary melts at 700$^o$C with glass composition. Se-rich ($x<$ 10\%) melts are fragile (${\cal M}>$40) and we note that the viscosity is quite low as it is estimated at 0.05 Pa.s at the target temperature. On the other hand, melts at $x>$ 24\% that are viewed as strong (${\cal M}\simeq$15-20), the viscosity increases to $\simeq$~200~Pa.s. 
In correspondence with the strong melts in the fragility index profile, a plateau is observed in the viscosity profiles which suggests that diffusion of the corresponding melts would be inhibited, as evidenced by the delayed homogenization. The delayed diffusion would naturally lead to a density-dependent organization of the vertical melt, leading to the melt fractionation shown, for example, in Figure \ref{fig2}. The high viscosities involved would hinder mixing, even when subject to rapid agitation. These striking increase in melt-viscosity will translate in correspondingly lowering of diffusivity D($x$), given that D depends inversely on melt viscosity through e.g. the Eyring relation\cite{eyring}.
\par
One thus expects a  melt at $x$ = 24\% weighted from an appropriate mix of Se glass ($x$=0\%) and As$_2$Se$_3$ glass ($x$=40\%), to rapidly (12 hours) alloy in the initial stages as Se diffuses into the 40\% glass and lowers the average stoichiometry to the 15\% $< x <$ 35\% range. At this point, the alloying process slows down qualitatively since the  strong melts formed now diffuse rather slowly. The decrease (respectively increase) of $D$ ($\eta$) by 4 orders of magnitude for melts in the 15\% $< x <$ 35\% composition range slows down the melt- mixing and thus homogenization as evidenced by the Raman profile data of Fig. \ref{Raman_prof} and \ref{fig2}. Since the melt alloying process is largely bottlenecked by diffusive processes, melt-rocking can only play a minor role in homogenization, a situation that has been also evidenced for the case of Ge-Se glasses\cite{r13,rajat}.
\par
A corollary to this observation is that melts once reacted for longer periods ($>$ 3 days) have progressively lower fragilities, with the lowest ‘{\em true}’ fragility measured eventually for the homogenized material. Agitation in a rocking furnace would essentially spread the fragile melts along the melt column without aiding the diffusion process necessary for equilibration at the weighed composition. It then follows that compositions with strong melts exist in a background of fragile melts. Melts at any arbitrary location would then, on average, have a higher fragility. It follows quite naturally that larger quantities of melts ($>$5 grams) would have a greater quantity of fragile melts, leading to higher fragility indices when measured at any location along the melt column. In other words, a mapping of fragilities along the column of underreacted melts would produce a high average fragility. A perusal of the sample synthesis methods of several groups\cite{r22,r23,feltz} shows that typical sample sizes of 20~grams were reacted at 750$^o$C while rocking for 20 hours. Based on the observations stated above, this length of time is insufficient for homogenizing melts of such large quantities. Since vastly different degrees of homogenization is achieved in the first 48 hours of homogenization, measured physical properties are likely to be a function of the reaction time and would differ with the synthesis times and sample sizes.
\par
\begin{figure}
\includegraphics*[width=0.9\linewidth]{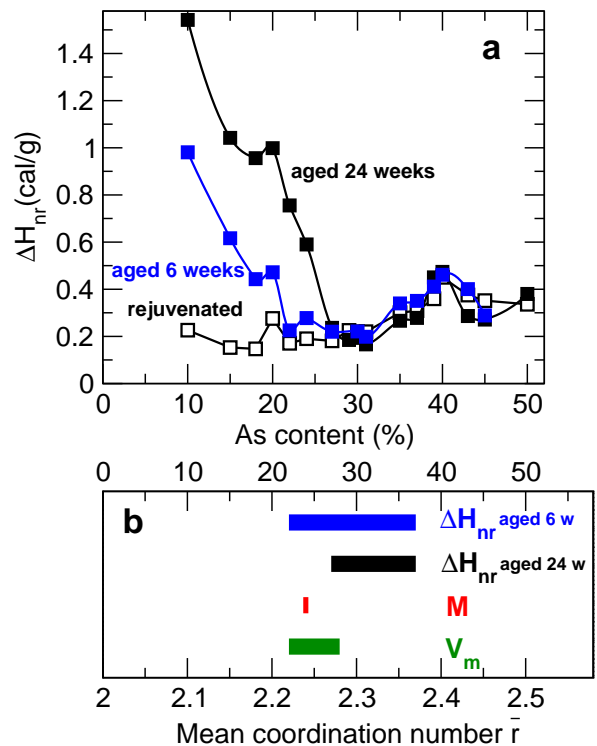}
\caption{\label{correl} a) Compositional variation of the non-reversing heat flow $\Delta$H$_{nr}$ at the glass transition after six weeks ($\color{blue}\blacksquare$), after 24 weeks of ageing at room temperature ($\blacksquare$), and after rejuvenation ($\square$). b) Comparison of the different determined compositional windows as a function of As content, and as a function of the mean coordination number $\bar r$.}
\end{figure}

\subsection{Calorimetric and fragility anomalies}
As highlighted in Figs. \ref{fragile}, \ref{activ}, and \ref{volume}, trends in fragility, activation energy, and molar volume exhibit an anomaly with As content for compositions roughly found between 22 and 28\% As, which result in a minimum value for ${\cal M}$, E$_a$ and V$_m$. These findings can be compared to calorimetric measurements that have been performed from a MDSC scan in the glass transition region, and which lead to the non-reversing heat flow $\Delta H_{nr}$, a quantity that captures most of the enthalpy of kinetic events associated with the slowing down of the relaxation. Interestingly, $\Delta H_{nr}$ displays a minimum as well in As-Se glasses (Fig. \ref{correl}a, similarly to previous findings\cite{r21}, and to other chalcogenides\cite{r13,r14,r15,r16,r17b,r17t}. The nearly vanishing of $\Delta H_{nr}$ which defines a reversibility window, is usually correlated with the existence of an adaptative intermediate phase\cite{thorpe,mmi,barre} that is found between the chalcogen-rich flexible phase, and the As-rich stressed rigid phase. The existence the reversibility window is, in fact, a direct consequence of the formation of an isostatic (stress-free) glass network that minimizes both stress (present at high As content) and the floppy modes (present at low As content) which would serve as impetus for relaxation. Without stress and without floppy modes, there is a significantly reduced thermodynamic driving force for relaxation and also a greater kinetic barrier to be overcome to induce relaxation. This direct relationship has been established from a Keating-Kirkwood model\cite{r17} showing that for an amorphous solid in which atoms are constrained by a potential containing both the stretching and bending interactions, the enthalpic overshoot in the glass transition endotherm is large for both flexible and stressed rigid networks, but it is minuscule for isostatic networks. It has also been demonstrated from pressure Raman experiments\cite{feng} showing that when $\Delta$ H$_{nr}\simeq$ 0, networks are stress-free and the Raman lines highly sensitive to small pressure changes, similarly to the well-known behavior in crystals\cite{crystals}.
\par
In Fig. \ref{correl}b, we represent the different compositional windows obtained from the measurements, and it reveals an interesting correlation between the different quantities. The fragility minimum at 24\% As is indeed obviously tied to the compositional window found for the molar volume (22\%$<x<$28\%) and to the reversibility window in $\Delta$H$_{nr}$ for fresh samples (22\%$<x<$37\%). However, one should note that the exact boundaries can not be fully inferred as for the case of Ge-Se\cite{r13,r17t}. In particular, the boundaries of the reversibility window are found to evolve with ageing time, especially on the chalcogen-rich side ($x<$22\%) and lead to a window approximatively centred around 30\% As after 6 months of ageing. 
\par
When represented as a function of the mean coordination number $\bar r$, one obviously finds a slight shift of the location of these anomalies with respect to the mean-field flexible to rigid transition of $\bar r$=2.4. It has been suggested that there may be some structural features\cite{wagner} specific to the present As-Se or its sulfide analogue\cite{dieman} which increase the isostatic character of the network at lower connectedness\cite{r21} and bring the onset of rigidity to lower As content. There are at this stage no experimental indication or signatures for such isostatic local structures although numerical evidence has been found in deep supercooled liquids close to the glass transition\cite{jncs2013} and in glasses\cite{hosokawa}. However, since all typical features\cite{book} of flexible, intermediate, and stressed rigid phases are found as the As content is increased, molar volume minima\cite{bourgel}, reversibility windows\cite{r14,r15,r16,r17,r17b,r17t}, fragility anomaly\cite{r13}, one can safely assign the flexible phase to compositions having $x<$20\%, and the stressed rigid phase to glasses with $x>$37\% As. These boundaries actually also agree with recent ab initio simulations showing various structural and dynamic anomalies to occur $\simeq$30\% As\cite{prl2013}, in agreement with those found in rigidity transitions driven by pressure\cite{prlns2,elliott}.

\subsection{Anomalies in scaling laws}
Scaling laws have been widely used for the analysis of relaxation phenomena in supercooled liquids. Here we investigate for the As-Se the validity of a scaling law relating the fragility index to the glass transition temperature. There is indeed a conventional wisdom suggesting that fragility increases with the glass transition temperature\cite{r9} which implicitely underscores the fact that energy barriers for relaxation increase with increasing T$_g$.
\par
The derivation of this scaling law uses the definition of equ. (\ref{vft0}), and calculates, using equ. (\ref{frag}), the fragility and the activation energy as a function of the glass transition temperature:
\begin{eqnarray}
\label{cor1}
{\cal M}={\frac {AT_g}{(T_g-T_0)^2\ln{10}}}
\end{eqnarray}
and:
\begin{eqnarray}
\label{cor2}
E_a={\frac {AT_g^2}{(T_g-T_0)^2}}
\end{eqnarray}
As $T_g$ is of the same order as $T_0$, one will have ${\cal M}$ and $E_a$ which will scale with $T_g$ and $T_g^2$, respectively. Note that this can be also independently derived from an alternative form using the Williams-Landel-Ferry (WLF) function for viscosity\cite{r1}. 
\begin{figure}
\includegraphics*[width=0.8\linewidth]{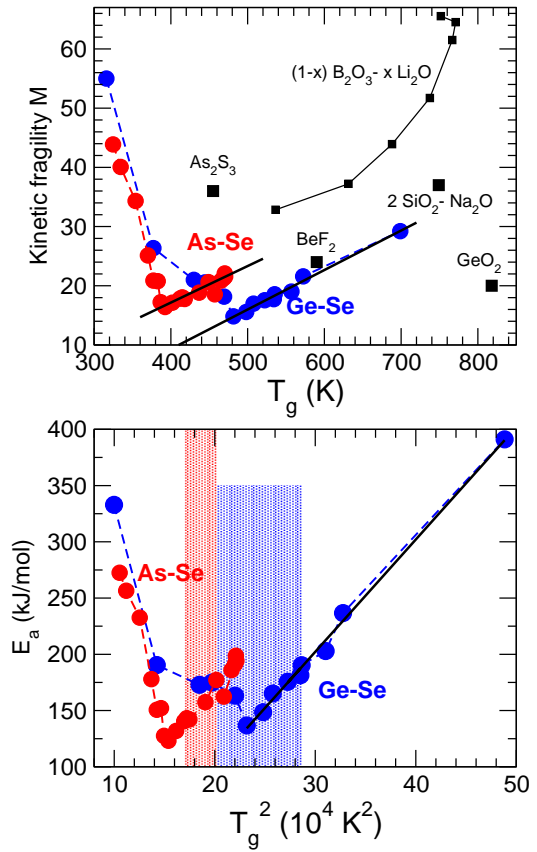}
\caption{\label{scaling} a) Fragility as a function of glass transition temperature in As-Se (red) and Ge-Se liquids (blue, Ref. \onlinecite{r13}), together with data for typical network glass formers \cite{r20,frag2,frag3} and binary glasses \cite{frag4}. b) Activation energy E$_a$ as a function of $T_g^2$. The shaded zones are the corresponding IPs\cite{r13}.
}
\end{figure}
\par
Using such scaling laws, Qin and McKenna\cite{qin} have shown that the correlations (\ref{cor1})-(\ref{cor2}) were fulfilled in a large class of hydrogen bonding organics, polymeric and metallic glass formers as all these systems show a linear increase of ${\cal M}$ with T$_g$, and $E_a$ with $T_g^2$. On the other hand, these authors remarked that inorganic glass formers did not seem to follow such scaling laws and ${\cal M}$ appeared to be nearly independent of the glass transition temperature. 
Figure \ref{scaling} represents the behavior of the fragility index (same as Fig. \ref{fragile}) as a function of the measured glass transition temperature T$_g$($x$) for the present As-Se, together with previous results on  Ge-Se \cite{r13}. It is found that when off-stoichiometric melts are followed in detail as a function of composition, one recoveres the scalings laws (\ref{cor1})-(\ref{cor2}) for only selected compositions corresponding merely to the stressed rigid and intermediate phase, i.e. one has a linear increase in ${\cal M}$($T_g$)for $x>$22\% in Ge-Se \cite{r13}, and for $x>$27\% in As-Se. A least-square fit to such compositions yields to ${\cal M}$=-7.53(5)+0.061(7)$T_g$ and to ${\cal M}$=-17.356+0.060(1)$T_g$ for As-Se and Ge-Se, respectively. The slope of both curves ($\simeq$ 0.06) is found to be somewhat lower than the one obtained \cite{qin} for metallic glass-formers (0.17), polymers (0.28) and hydrogen bonded liquids (0.25). However, as noted earlier\cite{qin}, no correlation is found among the various network formers such as B$_2$O$_3$\cite{frag4}, GeO$_2$\cite{r20} or As$_2$S$_3$\cite{frag2}. 
\par
Interestingly, a negative correlation is recovered, as for the case of Ge-Se\cite{r13}, which obviously can not be accounted from the VFT equation given that it would lead to unphysical behaviors such as the divergence of relaxation at a temperature $T_0>T_g$ or an increase of relaxation time $\tau$ with temperature\cite{r13}. An inspection of the different viscosity models shows that only the VFT equation (\ref{vft0}) (or (\ref{VFT})) can lead to a positive correlation (\ref{cor1}) between ${\cal M}$ and $T_g$. In fact, a simple Arrhenius behavior yields from equ. (\ref{vft0}) ${\cal M}$=$A/T_g\ln{10}$ and the MYEGA equation (\ref{MYEGA}) written in its compact form \cite{r5}:
\begin{eqnarray}
\label{myega0}
\log_{10}\eta=\log_{10}\eta_\infty+{\frac {K}{T}}\exp\biggl[{C/T}\biggr]
\end{eqnarray}
leads to:
\begin{eqnarray}
{\cal M}={\frac {K}{T_g}}\biggl(1+{\frac {C}{T_g}}\biggr)\exp\biggl[{C/T_g}\biggr]
\end{eqnarray} 
and for the apparent activation energy:
\begin{eqnarray}
E_a=K\ln{10}\biggl(1+{\frac {C}{T_g}}\biggr)\exp\biggl[{C/T_g}\biggr]
\end{eqnarray}
which both decrease as $T_g$ increases.
\par
\subsubsection{Constraints on the apparent activation energy}
Obviously, none of the forms for viscosity (equs. (\ref{vft0}) and (\ref{myega0}) can reproduce the obtained trend in ${\cal M}$(T$_g$) of Fig. \ref{scaling}. Rather than imagining an alternative form for $\eta$(T) able to display both positive and negative correlations in ${\cal M}$(T$_g$), we derive the conditions that should apply at the glass transition on the apparent activation energy $E$($T$) involved in a function describing viscosity (or relaxation or diffusion) with temperature:
\begin{eqnarray}
\label{newvisc}
\log_{10}\eta=\log_{10}\eta_\infty+{\frac {E(T)}{T}}
\end{eqnarray}
for which the fragility index is given by:
\begin{eqnarray}
\label{newvisc1}
{\cal M}={\frac {E(T_g)}{T_g}}-\biggl[{\frac {dA(T)}{dT}}\biggr]_{T=T_g}={\frac {E(T_g)}{T_g}}-E^{'}(T_g)
\end{eqnarray}
out of which we can calculate the slope:
\begin{eqnarray}
\label{slope}
{\frac {d{\cal M}}{dT_g}}=-E^{''}(T_g)+{\frac {E^{'}(T_g)}{T_g}}-{\frac {E(T_g)}{T_g^2}}
\end{eqnarray}
One must thus have for flexible glasses the conditions:
\begin{eqnarray}
E^{''}(T_g)+{\frac {E(T_g)}{T_g^2}}>{\frac {E^{'}(T_g)}{T_g}}
\end{eqnarray}
and:
\begin{eqnarray}
{\frac {d^2{\cal M}}{dT_g^2}}>0
\end{eqnarray}
It is easy to check that either equs. (\ref{vft0}) or (\ref{myega0}) only partially fulfill the differential inequalities so that they can not be used for the full trend in ${\cal M}$($T_g$) seen in Fig. \ref{scaling}. 
\subsubsection{Competing effects}
On a more general ground, it is seen that in order to find such negative correlations in ${\cal M}$(T$_g$), one must either have an extremum in ${\cal M}$ or in $T_g$, the former situation being found in the present chalcogenides whereas the latter is encountered for e.g., lithium borate glasses (Fig. \ref{scaling}) which show a steady increase of ${\cal M}$ with alkali concentration\cite{frag4} but a maximum in $T_g$. Both situations will lead to a branch with a negative slope in a (${\cal M}$, $T_g$) plot. Such a situation is not met in other glass-forming liquids where the positive correlation for ${\cal M}$(T$_g$) holds. By varying the molecular weight and cross-linking density in polymers, several authors\cite{auth1,auth2,auth3} have, indeed, demonstrated that such changes affect ${\cal M}$ and T$_g$ in a similar fashion, and an increase in both the fragility and T$_g$ as a function of e.g. the cross-link density.
\par
The physics which drives the trends in ${\cal M}$(T$_g$) actually results from two independent evolutions ${\cal M}$($x$) and T$_g$($x$) with composition, yet related to the underlying structural changes. It has been shown that $T_g$ is an accurate measure of network connectivity\cite{mmi1,mmi2,mmi3} and follows very precisely any coordination change induced by chemical alloying. This is not only true for chalcogenides but also for binary oxides such as alkali germanate or borate glasses for which a maximum in T$_g$ is obtained when the population of some higher-coordinated species maximizes with modifier content\cite{germanate,frag4,mauroborate}. 
Ultimately, it is seen that the change from a negative to a positive correlation is strongly tied to the minimum found for the fragility index ${\cal M}$ (Fig. \ref{correl}) and to the reversibility window. Based also on the previous findings on Ge-Se melts\cite{r13}, this seems to suggest that flexible chalcogenide glasses will generally display a negative correlation given that they always exhibit an increase of T$_g$ with connectivity but a decrease towards the fragility minimum found in the intermediate phase, whereas stressed rigid glasses will behave the opposite way. 
\par
On the other hand, flexible Se-rich supercooled liquids have the same structural morphology as cross-linked polymeric melts for which the VFT or the WLF equations fit quite accurately the viscosity evolution, and which subsequently lead to the reported positive correlation\cite{qin} for ${\cal M}$($T_g$). Furthermore, the glass transition temperature variation of such flexible As-Se glasses can be quite accurately described\cite{naumis_mmi} from the Gibbs-Di Marzio equation\cite{gibbs} derived for cross-linked polymers. When this category of glass-forming liquids is considered, it is therefore quite unexpected that glasses with a chain-like structure do not follow the positive correlation.

\section{\label{conclusion} Conclusion}
Melt homogenization is a critical subject that is, unfortunately, often overlooked in the literature. In this article, we have shown that for off-stoichiometric compositions important physical properties crucially depend on the melt processing, and the way melts homogenize at the reaction temperature. For the present case of As-Se liquids, it is found that while compositional trends (i.e. the relative variation) in fragility index ${\cal M}$ do not differ with samples having been homogenized over smaller reaction times\cite{r23}, the absolute magnitude of ${\cal M}$ is substantially altered with differences up to about a factor two. Carefully homogenized systems over long periods then lead to melts having one of the lowest known fragilities (${\cal M}\simeq$ 16.4 for the 24\% As glass), lower than the archetypal strong-glass former silica. This allows understanding the origin of the slow homogenization dynamics which arises from the nature of these {\em super-strong melts} in the range of 25-35\% As, having potentially large viscosities, and prevents from fast mixing. Raman profiling confirms the overall tendencies and clearly shows that an important spread in composition still exists along the melt column after 3 days alloying (Fig. \ref{fig2}), a time interval that already exceeds what has been routinely used in earlier studies. These results appear to be generic, and parallel to recent findings on another important chalcogenide system, Ge-Se\cite{r13}.
\par
Having obtained glasses and melts of proven homogeneity, we then investigate the effect of As composition on melt fragility. The obtained fragility minimum appears to be directly related to flexible to stressed rigid transitions, and to the reversibility window revealed by the nearly vanishing of the non-reversing heat flow $\Delta$H$_ {nr}$. It is also found that homogenization impacts glass network packing such as molar volumes which display a minimum for the same compositions, a feature that originates from the stress-free nature of the network backbone. In this respect, the present As-Se reproduces all the well-established features of isostatic windows such as space-filling tendencies\cite{bourgel}, thermally-reversing character of the glass transition\cite{r16}, and fragility anomalies\cite{r13}.
\par
Given the reported impact of glass sample make up\cite{r17b,r17t} on physical properties of network glasses, the crucial role played by melt homogenization and glass processing should once again be emphasized at the end. Based on the present work, and on the previous example of Ge-Se\cite{r13}, we are convinced that glasses of proven homogeneity will certainly exhibit the intrinsic behavior of adaptative networks undergoing a flexible to rigid transtion over a finite width of stress-free compositions.
\par
PB acknowledges support from NSF grant DMR-08-53957. MM acknowledges support from ANR grand 09-BLAN-0109-01, from the Franco-American Fulbright Commission and from International Materials Institute.

\end{document}